\documentclass[a4paper,prl,aps,twocolumn]{revtex4}
\usepackage{graphicx}
\usepackage{amsmath}
\usepackage{bm}
\usepackage{braket}
\usepackage{psfrag} 
\usepackage{latexsym} 
\usepackage{amstext}
\usepackage{amsxtra} 
\usepackage{float}
\usepackage{upgreek} 
\usepackage{amssymb}
\usepackage{times} 
\usepackage{pdfpages}
\usepackage[hidelinks]{hyperref}
\usepackage{pgffor}

\begin{document}

\newcommand{\kB}{k_{\rm B}}
\newcommand{\cs}{$\clubsuit \; $}
\newcommand{\downstate}{\left\vert\downarrow\right\rangle}  
\newcommand{\upstate}{\left\vert\uparrow\right\rangle}
\newcommand{\overbar}[1]{\mkern 1.5mu\overline{\mkern-1.5mu#1\mkern-1.5mu}\mkern 1.5mu}
\newcommand{\expect}[1]{\langle#1\rangle}

\title{
Two- and Three-body Contacts in the Unitary Bose Gas
}

\author{Richard J. Fletcher$^1$, Raphael Lopes$^1$, Jay Man$^1$, Nir Navon$^1$, Robert P. Smith$^1$,\\Martin W. Zwierlein$^2$, Zoran Hadzibabic$^1$}
\affiliation{$^1$Cavendish Laboratory, University of Cambridge, J.~J.~Thomson Avenue, Cambridge CB3~0HE, United Kingdom\\
$^2$Department of Physics, MIT-Harvard Center for Ultracold Atoms, and Research Laboratory of Electronics, MIT, Cambridge, Massachusetts 02139, USA
}

\begin{abstract}
In many-body systems governed by pairwise contact interactions, a wide range of observables is linked by a single  parameter, the two-body contact, which quantifies two-particle correlations.
This profound insight has transformed our understanding of strongly interacting Fermi gases.
Here, using Ramsey interferometry, we study coherent evolution of the resonantly interacting Bose gas, and show that it cannot be explained by only pairwise correlations. Our experiments reveal the crucial role of three-body correlations arising from Efimov physics, and provide a direct measurement of the associated three-body contact.
\end{abstract}

\date{\today}



\maketitle

A fundamental challenge in many-body quantum physics is to connect the macroscopic behaviour of a system to the microscopic interactions between its constituents. 
In ultracold atomic gases the strength of interactions is most commonly characterised by the $s$-wave scattering length $a$,  which can be tuned via Feshbach resonances~\cite{Chin:2010}.
On resonance $a$ diverges and one reaches the unitary regime, in which the interactions are as strong as allowed by quantum mechanics.
This regime has been extensively studied in Fermi gases~\cite{Inguscio:2008,Zwerger:2011,Zwierlein:2014}, while the unitary Bose gas represents a new experimental frontier~\cite{Rem:2013,Fletcher:2013,Makotyn:2014,Eismann:2016,Hu:2016,Jorgensen:2016}.

In these systems, universal properties of the short-range particle correlations imply universal thermodynamic relations between macroscopic observables such as the momentum distribution, energy, and the spectroscopic response~\cite{Tan:2008a,Punk:2007,Baym:2007,Braaten:2010,Schneider:2010,Castin:2010,Braaten:2011,Werner:2012,Smith:2014}.
In the case of (mass-balanced) two-component Fermi gases, at the heart of these relations is a single fundamental thermodynamic parameter, the two-body contact density $C_2$, which measures the strength of two-particle correlations.
However, the case of the Bose gas is more subtle. In this system Efimov physics gives rise to three-body bound states~\cite{Efimov:1970,Kraemer:2006,Zaccanti:2009,Ferlaino:2011,Wild:2012,Machtey:2012,Roy:2013}, and more generally introduces three-particle correlations that cannot be deduced from the knowledge of pairwise ones~\cite{Braaten:2011,Werner:2012,Smith:2014,Piatecki:2014}. 
The implication for many-body physics is that complete understanding of the macroscopic coherent phenomena requires knowledge of both $C_2$ and its three-body analogue $C_3$~\cite{Braaten:2011,Werner:2012,Smith:2014}.

The relative importance of three-particle correlations generally grows with the strength of interactions. At moderate interaction strengths $C_2$ was measured spectroscopically, but $C_3$ was not observed~\cite{Wild:2012}. However, the momentum distribution of the unitary Bose gas~\cite{Makotyn:2014} suggested deviations from two-body physics~\cite{Smith:2014, Barth:2015}.

\begin{figure}[tbp] 
\includegraphics[width=1\columnwidth]{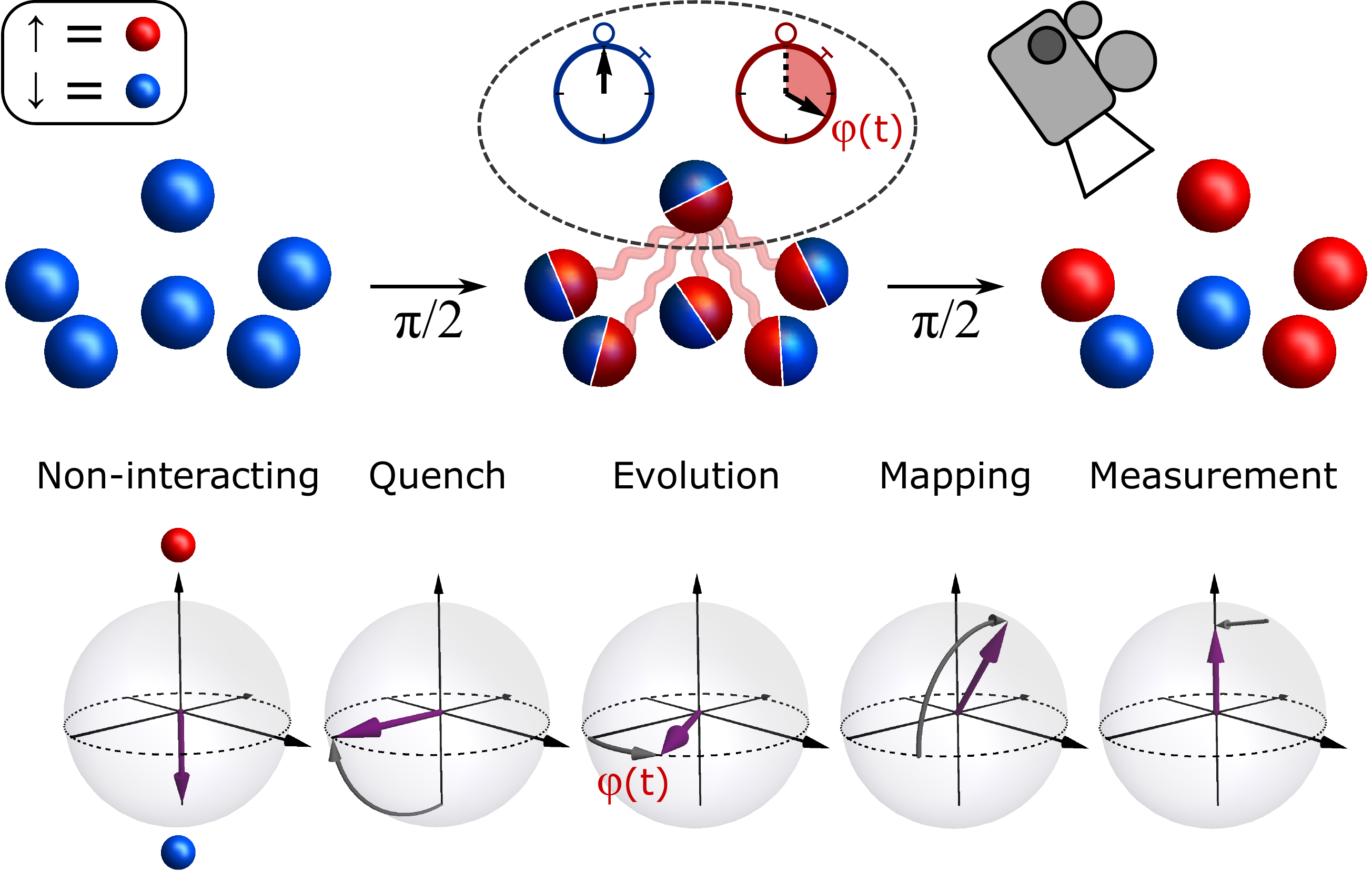}
\caption{
\textbf{Ramsey interferometry of a many-body system.} The first $\pi/2$ pulse puts each atom in a superposition of $\uparrow$ (red) and $\downarrow$ (blue) states. Strong interactions between the red components cause the relative phase of the superposition to advance by $\varphi$. The second $\pi/2$ pulse maps $\varphi$ onto spin polarization, which is measured by absorption imaging. Below, the stages of our protocol are illustrated in terms of the collective spin on the Bloch sphere.
}
\label{fig:fig1}
\end{figure}

Here we interferometrically measure both $C_2$ and $C_3$ in a resonantly interacting thermal Bose gas, and find excellent agreement with theoretical predictions.
The idea of our experiment is illustrated in Fig.~\ref{fig:fig1}. We perform radio-frequency (RF) Ramsey interferometry on a gas of atoms with two internal (spin) states, $\uparrow$ and $\downarrow$, and use a magnetic Feshbach resonance to enhance $\uparrow\uparrow$ interactions, while both $\uparrow\downarrow$ and $\downarrow\downarrow$ interactions are negligible.
For a measurement at a given magnetic field, we initially prepare a gas in the $\downarrow$ state, and then use an RF pulse to put each atom into an equal superposition of $\uparrow$ and $\downarrow$. This corresponds to an interaction quench that initiates many-body dynamics. Focusing on one particular atom, during the subsequent evolution its $\uparrow$ component accumulates a phase $\varphi$ due to interactions with the other $\uparrow$ components in the surrounding cloud. As we formally show in the Supplementary Materials, the rate at which $\varphi$ accumulates reflects many-body correlations that would develop in a purely-$\uparrow$ system with half the total density. Meanwhile, the $\downarrow$ component serves as a non-interacting phase reference, which allows us to read out $\varphi$ interferometrically~\cite{rudiNote}. This is accomplished by a second RF pulse, which maps $\varphi$ onto a spin-population imbalance that we measure directly.

In Fig.~\ref{fig:fig1} our protocol is also shown on the Bloch sphere, in terms of the collective spin $\vec{S}$. During the evolution of the equal-superposition state, $\vec{S}$ precesses in the equatorial plane at a rate $\Omega\equiv \dot{\varphi}$. In the Supplementary Materials we derive the relationship between $\Omega$ and the two- and three-body contacts:
\begin{equation}
\Omega=\frac{\hbar}{4\pi m}\left(\frac{1}{ n a} C_2 +  \frac{5.0\,\pi^2}{n} C_3 \right) , 
\label{eqn:shiftContact}
\end{equation}
where $m$ is the atom mass, $n$ the density of the $\uparrow$ component, and $a$ the $\uparrow\uparrow$ scattering length. 
Away from unitarity, $C_2 \sim n^2a^2$ and $C_3\sim n^3 a^4$~\cite{Werner:2012,Smith:2014}, and the ratio of the $C_3$ and $C_2$ contributions to $\Omega$ is ${\sim}\,n|a|^3 \ll 1$.
At unitarity, both contacts saturate at their maximal values; in a thermal gas $C_2\sim n^2\lambda^2$ and $C_3\sim n^3\lambda^4$, where $\lambda$ is the thermal wavelength. 
The crucial advantage of using the precession of the Bloch vector to observe three-particle correlations is that the $C_2$ contribution to $\Omega$ vanishes at unitarity (where $|a|\rightarrow \infty$).

\begin{figure}[tbp] 
\includegraphics[width=1\columnwidth]{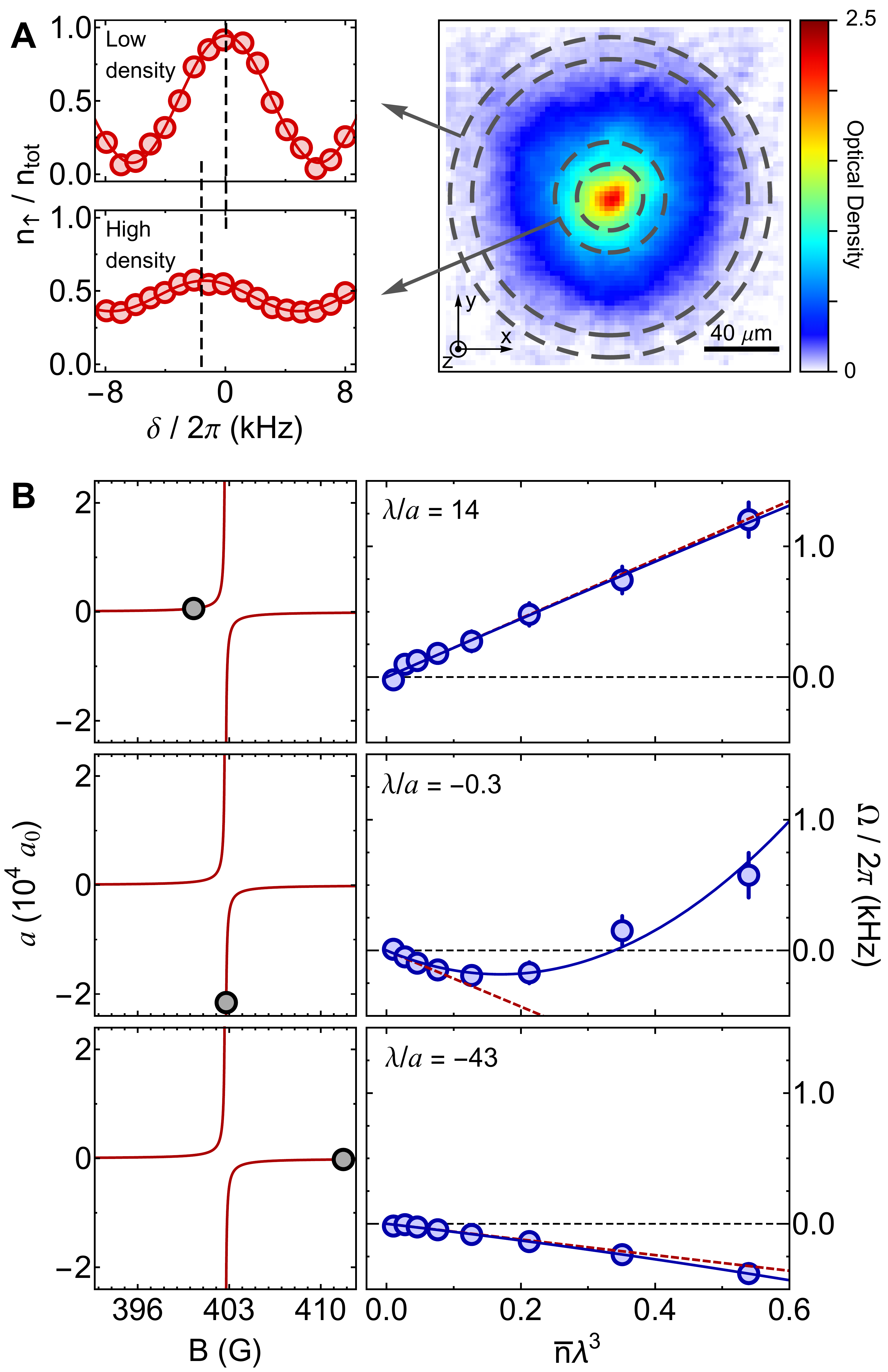}
\caption{ \textbf{Density-dependent phase winding.} \textbf{(A)} Ramsey oscillations of the spin-$\uparrow$ density as a function of the RF detuning $\delta$. Oscillations at different positions in the trap reveal the density dependence of $\varphi$. Strong interactions both shift the Ramsey fringes and reduce their contrast.
\textbf{(B)} For weak interactions (top and bottom) $\Omega$ varies linearly with density, but close to unitarity (middle) it shows non-linear behaviour that reveals the influence of three-body physics. At all scattering lengths the data are fitted well by a second-order polynomial (solid blue lines); the red dashed lines show the linear parts of the fits. All error bars show standard fitting errors.}
\label{fig:fig2}
\end{figure}

Our experimental setup is described in Ref.~\cite{Campbell:2010}. 
We work with $^{39}$K atoms prepared in an optical harmonic trap with frequencies $(\omega_x,\omega_y,\omega_z)/2\pi=(48.5,56.5,785)$~Hz. Our two spin states, labelled in the low-field basis, are $\upstate\equiv\ket{F=1,m_F=1}$ and $\downstate\equiv\ket{F=1,m_F=0}$.
We tune the $\uparrow\uparrow$ scattering length $a$ using a Feshbach resonance centred on $B_0 = 402.70(3)$~G~\cite{SI}. In all our experiments $|a|>300~a_0$ while the moduli of the $\uparrow\downarrow$ and $\downarrow\downarrow$ scattering lengths are $< 10~a_0$~\cite{Lysebo:2010}, where $a_0$ is the Bohr radius.
Near $B_0$ the bare splitting of the $\uparrow$ and $\downarrow$ states is ${\approx}\, 99$~MHz.
We prepare clouds at the critical point for Bose-Einstein condensation, with a phase-space density $n_{\rm tot} \lambda^3\sim 2.6$ at the trap centre, where $n_{\rm tot}$ is the number density and the cloud temperature of $370$~nK corresponds to $\lambda\approx 8600~a_0$.
The duration of each $\pi/2$ pulse is $t_{\rm p}=17~\mu$s, and the evolution time between the pulses, $T$, is varied up to $130~\mu$s. 
At the end of the whole Ramsey sequence we measure the fractional $\uparrow$ population, $n_\uparrow/n_{\rm tot}$, by {\it in situ} absorption imaging along $\hat{z}$ (see Fig.~\ref{fig:fig2}A).
In Bose gases strong coherent interactions are generally accompanied by significant inelastic losses, but on the timescale of our experiments the atom loss at our highest density is $<10\%$.

To measure the density-dependent $\Omega$ we scan the detuning of the RF source from the non-interacting resonance, observe Ramsey oscillations of the spin populations, and extract the detuning, $\delta_0$, for which $n_\uparrow/n_{\rm tot}$ is maximal (see Fig.~\ref{fig:fig2}A). 
We exploit the fact that the atoms are essentially stationary during the Ramsey sequence to simultaneously extract $\delta_0$ for a wide range of densities, from the local oscillations of $n_\uparrow/n_{\rm tot}$ in different regions of the cloud.
Most generally
\begin{equation}
\delta_0=-\frac{\varphi+\Delta\varphi_{\rm p}}{T+ 4t_{\rm p}/\pi}, 
\label{eqn:fringeCentre}
\end{equation}
where $\Delta\varphi_p$ is any interaction-induced phase accumulated during the RF pulses~\cite{SI}.
For constant $\Omega$ (so $\varphi=\Omega T$) and $T\gg t_{\rm p}$, Eq.~(\ref{eqn:fringeCentre}) reduces to the intuitive $\delta_0 = - \Omega$.
For measurements at low density and away from unitarity this is an excellent approximation. For more accurate studies at high densities, or close to unitarity, we perform differential measurements, in which we extract $\delta_0$ for various evolution times and reconstruct the instantaneous $\Omega(t)$. This mitigates the small effects of the non-zero pulse duration and also allows us to study the dynamics of $C_2$ and $C_3$.

In Fig.~\ref{fig:fig2}B, we show the density dependence of $\Omega$ (assuming for now $\Omega = - \delta_0$) for weak and nearly-unitary interactions. 
Here $\overbar{n}$ is the $\uparrow$-density experienced by an atom, averaged over the imaging line-of-sight and a small radial bin in the image plane.
For weak interactions $\Omega\propto \overbar{n}$, consistent with the expected dominance of two-body correlations for $n|a|^3 \ll 1$. 
However, close to unitarity $\Omega(\overbar{n})$ is non-linear and even changes sign, which cannot be explained by two-body physics.

For a quantitative analysis we first focus on very low densities. In this limit $\Omega$ is dominated by two-body correlations at all interaction strengths.
From the measurements of  $\Omega(\overbar{n})$ we extract the initial slope $\alpha \equiv \partial\Omega/\partial\overbar{n}|_{\overbar{n}=0}$ (see Fig.~\ref{fig:fig2}B), which gives the behaviour of $\Omega$ at vanishing density.

\begin{figure}[tbp] 
\includegraphics[width=1\columnwidth]{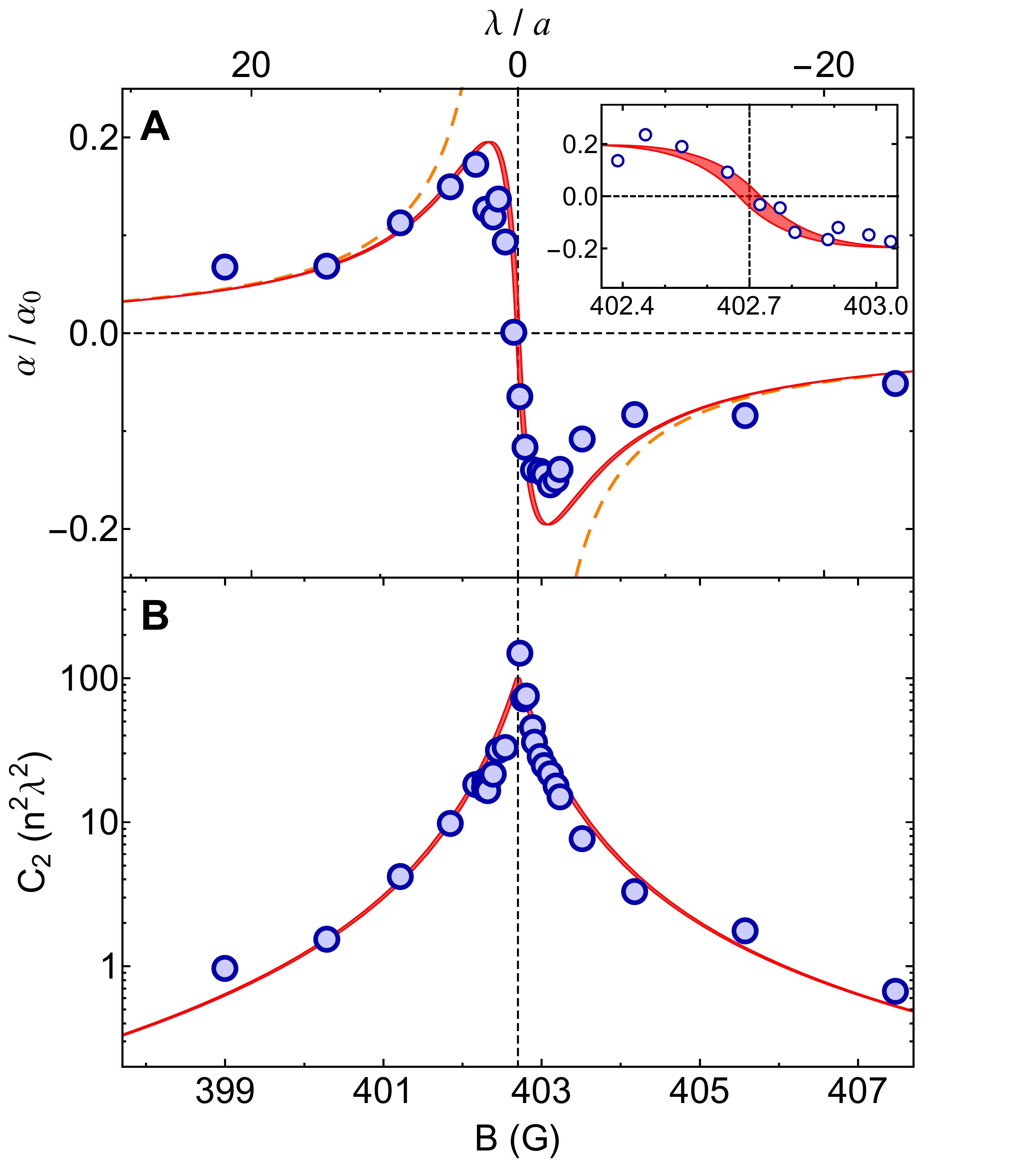}
\caption{ \textbf{Two-body contact.} \textbf{(A)} Initial slope $\alpha$ of $\Omega(\overbar{n})$, normalised to $\alpha_0 = 8\pi\hbar\lambda/m$. The solid red line shows the theoretical prediction~\cite{Kokkelmans:1997_b}, and the dashed orange line its weakly-interacting limit, $\alpha/\alpha_0=a/\lambda$. Inset: measurements close to the resonance (see text).  \textbf{(B)} $C_2$ extracted from $\alpha$. The red line is the theoretical prediction of~\cite{Braaten:2013}. The thickness of the red lines reflects the uncertainty in $B_0$. 
}
\label{fig:fig3}
\end{figure}

In Fig.~\ref{fig:fig3}A we plot $\alpha$ across the Feshbach resonance, for $T=125~\mu$s. 
The solid red line shows $\alpha=8\pi\hbar\overbar{\Re(f)}/m$, where $\overbar{\Re (f)}$ is the real part of the scattering amplitude $f$~\cite{Kokkelmans:1997_b}, averaged over the thermal momentum distribution; the dashed orange line is the weakly-interacting limit $\alpha = 8 \pi\hbar a /m$.
Using Eq.~(\ref{eqn:shiftContact}), from our measurements we extract $C_2/n^2=\alpha 4\pi m a /\hbar$. This is plotted in Fig.~\ref{fig:fig3}B, along with an analytic prediction for $C_2$~\cite{Braaten:2013}. Over two orders of magnitude in $C_2$ we find excellent agreement between theory and our data.

In our search for $C_3$, a key prediction of Eq.~(\ref{eqn:shiftContact}) is that the $C_2$ contribution to $\Omega$ vanishes exactly at $B_0$. In the inset of Fig.~\ref{fig:fig3} we show measurements focused on the resonance region and verify that this is indeed the case. Here,
we measure $\delta_0$ for two evolution times, $T_1=40~\mu$s and $T_2=125~\mu$s, to assess the instantaneous $\Omega$ at $t=82.5~\mu$s according to Eq.~(\ref{eqn:fringeCentre}).
We also varied $T_1$ and $T_2$ and found that $\alpha$ is always consistent with the equilibrium theory curve (red shading). This is in agreement with our simulations of the two-particle dynamics after an interaction quench~\cite{SI}. We theoretically find that $C_2$ equilibrates on a timescale $\tau_2$ which is ${\sim}\,ma^2/\hbar$ away from the Feshbach resonance and ${\sim}\,m\lambda^2/\hbar$ at unitarity; for our experimental parameters $\tau_2$ is shorter than the first RF pulse.

\begin{figure*}[tbp] 
\includegraphics[width=\textwidth]{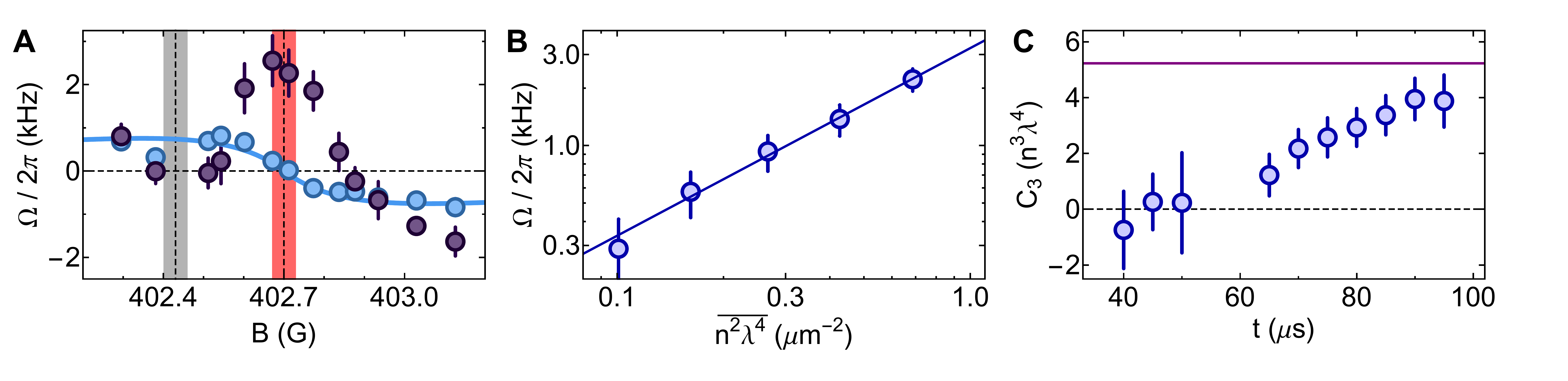}
\caption{\textbf{Three-body contact.} 
\textbf{(A)} $\Omega$ at $t=90~\mu$s, for densities $\overbar{n}\lambda^3=0.13$ (light blue) and $\overbar{n}\lambda^3=0.54$ (dark blue). 
For reference we also show the two-body prediction for the lower-density data (blue line). 
The red band indicates the position of the Feshbach resonance, and the grey band the location of the previously observed minimum in the three-body loss rate~\cite{Zaccanti:2009}.
\textbf{(B)} Density dependence of $\Omega(B_0)$; here $t=90~\mu$s data is averaged within the red band shown in \textbf{A}. The linear fit (blue line) gives a slope $1.0(1)$, in excellent agreement with the three-body scaling law. \textbf{(C)} The three-body contact density $C_3$ as a function of time after the interaction quench. The horizontal purple line shows the theoretical prediction for the equilibrium unitary Bose gas.}
\label{fig:fig4}
\end{figure*}
We now turn to higher densities and strong interactions, where the effect of $C_3$ should be prominent. We always reconstruct the instantaneous $\Omega(t)$, and in Fig.~\ref{fig:fig4}A we show it for $t = 90~\mu$s and two different densities. 
At high density we clearly observe a non-zero $\Omega$ at unitarity, which as per Eq.~(\ref{eqn:shiftContact}) cannot arise from a $C_2$ contribution (see also~\cite{Li:2012, Stoof:2014}).  
Additionally, away from unitarity, at $B<B_0$, we see an intriguing suppression of $\Omega$, which coincides with the previously observed strong suppression of three-body losses (at $a \sim 5600\, a_0$)~\cite{Zaccanti:2009}.

Here we focus on the non-zero $\Omega$ at unitarity, and verify that it arises from three-particle correlations, by looking at its scaling with density. A $C_3$ contribution to $\Omega$ should scale as $n^2\lambda^4$. 
In Fig.~\ref{fig:fig4}B we show that on a log-log plot  $\Omega(B_0)$ clearly shows linear dependence on $\overbar{n^2\lambda^4}$~\cite{SI}. The fitted slope is $1.0(1)$, in excellent agreement with the three-body scaling law.

Finally, we study the magnitude of the unitary $C_3$. In contrast to $C_2$, we observe a gradual development of $C_3$ over the timescale of our experiment (see Fig.~\ref{fig:fig4}C), which means that after the interaction quench the three-body correlations develop slower than the two-body ones. For $t\lesssim 50~\mu$s the three-body contact is consistent with zero (within our error bars), while at our longest times, $t\approx 100~\mu$s, it approaches the theoretical expectation for the equilibrium unitary gas,  $C_3/(n^3\lambda^4) \approx 5.2$~\cite{Smith:2014}.

Our measurements provide the first conclusive observation of the effects of three-body correlations on the coherent behaviour of a many-body system.
The non-equilibrium dynamics of the three-body contact is an interesting open problem for future study.
It would be very exciting to extend our technique to a deeply-degenerate gas, for which $C_3$ is not even theoretically known~\cite{Smith:2014}. In our harmonic-trap setup, starting with a non-interacting Bose condensate would result in prohibitively short lifetimes after the quench to unitarity, but this problem could be mitigated by using a uniform trapping potential~\cite{Gaunt:2013}.

We are indebted to Eric Braaten for his crucial input towards the derivation of Eq.~(1), and critical reading of the manuscript. We thank Martin Robert de Saint Vincent for contributions in the early stages of the project, Maximilian Sohmen for experimental assistance, and Servaas Kokkelmans, Eric Cornell, David Papoular, F\'{e}lix Werner, Isabelle Bouchoule, Isaac Chuang and Jean Dalibard for helpful discussions. This work was supported by EPSRC [Grant No. EP/N011759/1], ERC (QBox), ARO and AFOSR. N.N. acknowledges support from Trinity College, Cambridge, R.P.S. from the Royal Society and R.L. from the E.U. Marie-Curie program [Grant No. MSCA-IF-2015 704832].

\bibliographystyle{Science}

\clearpage


\section*{Supplementary material}

# Two- and Three-body Contacts in the Unitary Bose Gas

## Supplementary Materials

Richard J. Fletcher[1], Raphael Lopes[1], Jay Man[1], Nir Navon[1], Robert P. Smith[1],
Martin W. Zwierlein[2], and Zoran Hadzibabic[1]

[1]*Cavendish Laboratory, University of Cambridge, J. J. Thomson Avenue, Cambridge CB3 0HE, United Kingdom*
[2]*Department of Physics, MIT-Harvard Center for Ultracold Atoms,
and Research Laboratory of Electronics, MIT, Cambridge, Massachusetts 02139, USA*

**EXPERIMENTAL CALIBRATIONS**

**Calibration of the Feshbach resonance position**

The hyperfine state $|F = 1, m_F = 1\rangle$ of $^{39}$K has a broad Feshbach resonance located at $\sim 400$ G. In Table S1 we summarise previous theoretical predictions and experimental measurements for the resonance position $B_0$.

| Source | $B_0$ (Theoretical) | $B_0$ (Experimental) |
|---|---|---|
| [1] | 402.4(2) G | 403.4(7) G |
| | | 401.5(5) G |
| [2] | | 402.50(3) G |
| [3] | 402.9 G | |
| [4] | 402.4(2) G | 402.6(2) G |

**TABLE S1. Previous values for the Feshbach resonance location.**

Confirming that the two-body contribution to $\Omega$ vanishes exactly at unitarity requires a precise knowledge of the resonance position, and the spread of values and the uncertainties in Table S1 are too large for our purposes. We therefore obtain a new calibration of the resonance position, using the anisotropic expansion of a dilute thermal gas from an anisotropic trap. A strongly-interacting gas displays hydrodynamic flow that entails transfer of energy from one expansion axis to another, through two-body elastic scattering. For two identical bosons with relative wavevector $k$, the scattering cross section

$$\sigma = \frac{8\pi a^2}{1 + k^2 a^2}, \quad (S1)$$

is maximal at the Feshbach resonance where $a = \infty$.

For this calibration we initially prepare an essentially non-interacting thermal gas of $|F = 1, m_F = 0\rangle$ atoms at a temperature of 580 nK, in a trap with frequencies $(\omega_x, \omega_y, \omega_z)/(2\pi) = (48.5, 56.5, 1080)$ Hz. We fix the Feshbach field $B$ to a particular (variable) value, then switch off the trapping potential and simultaneously perform an RF $\pi$-pulse of duration 34 $\mu$s that transfers all the atoms to the interacting $|1, 1\rangle$ state. After 13 ms of time-of-flight expansion, we take an absorption image of the gas along the horizontal axis $\hat{y}$, and extract the aspect ratio $\sigma_z/\sigma_x$ from the fitted Gaussian widths $\sigma_z$ and $\sigma_x$. In these measurements the average phase

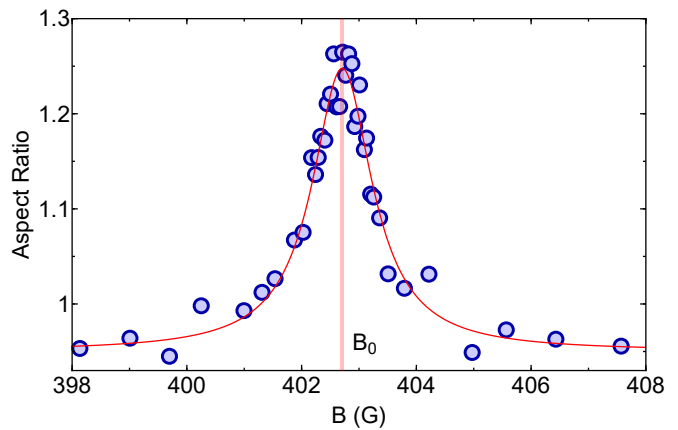

FIG. S1. **Calibration of the Feshbach resonance.** The cloud aspect ratio measured after 13 ms of time-of-flight, as a function of the Feshbach field $B$. The red line shows a Lorentzian fit, with the central value and its uncertainty indicated by the red vertical band. The aspect ratio is slightly below unity for weak interactions due to the non-infinite time-of-flight; for an ideal gas the fit gives 0.95, while we calculate 0.97.

space density is $\langle n_{tot}\lambda^3 \rangle = 0.03$, where $n_{tot}$ is the number density and $\lambda = 368$ nm is the thermal wavelength, and we do not observe any three-body losses during the expansion.

The aspect ratio is plotted in Fig. S1 as a function of $B$. The data is well-fitted by a Lorentzian function which we use to extract $B_0 = 402.70(3)$ G; the result remains within the error bar if we use a Gaussian fit instead.

**Residual losses and heating from three-body recombination**

After the initial quench to strong interactions, three-body recombination leads to a loss of atoms and heating of the cloud. In the majority of the main paper, densities are labelled by their initial values. However, in Fig. 4B for a careful analysis of the scaling of $\Omega$ at unitarity with the local average value of $n^2\lambda^4$, we take these small losses into account. We define the correction parameter $\chi(t)$ by

$$\overline{n^2\lambda^4}(t) = \chi(t) \, \overline{n_0^2 \lambda_0^4} \quad (S2)$$

where $n_0$ is the initial density, and $\lambda_0$ is the thermal wavelength corresponding to the initial temperature $T_0$. The bar

denotes the average value experienced by an atom, which is obtained by averaging the quantity $n \times (n^2 \lambda^4)$ over space.

The loss equation for the density $n$ of a thermal Bose gas at unitarity is [5, 6]

$$\dot{n}(t) = -L_3(t) \frac{T_0^2}{T^2(t)} n^3(t). \tag{S3}$$

Note that we have permitted a time-dependence of the loss coefficient $L_3$. We find that the three-body contact density $C_3$ gradually develops after the interaction quench, and since $L_3 \propto C_3$ [7, 8] it should not be time-independent.

At unitarity, atoms with low momentum are preferentially lost, leading to heating of the gas. This is captured by the scaling law $nT^\beta = n_0 T_0^\beta$ where $\beta = 18/13$ [5, 6]. Substitution of this scaling law leads to the following equation for $n(t)$,

$$n(t) = \left[ (1-\nu)A(t) + n_0^{1-\nu} \right]^{\frac{1}{1-\nu}}, \tag{S4}$$

where $\nu = 3 + 2/\beta \approx 4.2$ and $A(t)$ encodes all the time-dependence of the losses. For a particular initial density profile $n_0(z)$ along the imaging axis, the surviving fraction of the column density, $\zeta(t)$, and the correction parameter defined in Eq. (S2), $\chi(t)$, are given by

$$\zeta(t) = \frac{\int_{-\infty}^{\infty} dz \, n(z,t)}{\int_{-\infty}^{\infty} dz \, n_0(z)}, \tag{S5}$$

$$\chi(t) = \frac{\int_{-\infty}^{\infty} dz \, [n(z,t)]^\nu [n_0(z)]^{3-\nu}}{\int_{-\infty}^{\infty} dz \, n_0^3(z)} \frac{1}{\zeta(t)}. \tag{S6}$$

Both quantities are single-valued parametric functions of $A(t)$; for a particular density profile $n_0(z)$, measurement of the integrated surviving fraction $\zeta$ therefore fully determines $\chi$. We numerically construct the function $\chi(\zeta)$ from Eqs. (S5-S6) assuming a Gaussian $n_0(z)$.

At each radius from the cloud centre, we obtain the correction function $\chi(t)$ as follows. Starting with the same non-interacting cloud as for our Ramsey protocol, we apply an RF $\pi/2$-pulse, then wait a variable time $t$ before measuring the column density of the interacting $|\uparrow\rangle \equiv |F = 1, m_F = 1\rangle$ component by state-selective absorption imaging. We then use our numerical function $\chi(\zeta)$ to convert the measured $\zeta(t)$ curves into $\chi(t)$, which is applied to the data according to Eq. (S2).

In Fig. S2 we show a log-log plot of $\Omega$ at unitarity for $t = 90 \, \mu s$, as a function of $\overline{n^2 \lambda^4}$. The grey points denote the uncorrected data, and the blue points the data after applying Eq. (S2). The corrected data shows a clear linear dependence, with a fitted slope of 1.0(1), which is in excellent agreement with the theoretical expectation of unity. The uncorrected data shows the same slope for low densities, but curves as the density, and hence losses, increases.

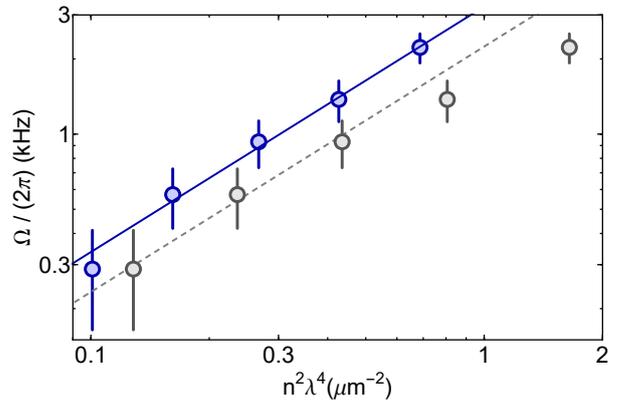

FIG. S2. **Effect of losses and heating.** $\Omega$ at unitarity for $t = 90 \, \mu s$, against $\overline{n^2 \lambda^4}$. The blue points are the data corrected according to Eq. (S2), and the blue line is the corresponding linear fit, with slope 1.0(1). The grey points show the uncorrected data; the slope of the dashed grey line is fixed to unity.

## RAMSEY SPECTROSCOPY OF MANY-PARTICLE SYSTEMS

In our experiment we perform spectroscopy on a strongly-interacting Bose gas, in which two hyperfine levels constitute an effective spin-1/2 system $\{\uparrow, \downarrow\}$. The total spin of the $N$-particle gas is given by $\hat{\vec{S}} = (\hat{S}_x, \hat{S}_y, \hat{S}_z)$, where $\hat{S}_m = \sum_{j=1}^{N} \hat{s}_{j,m}$ is a collective spin component, and $\hat{s}_{j,m}$ gives the spin of the $j^{\text{th}}$ particle along $m \in \{x, y, z\}$.

Our Ramsey protocol probes the collective Bloch vector, defined by $\vec{S} = \langle \hat{\vec{S}} \rangle$. Each atom in our (thermal) gas is initially in the non-interacting $\downarrow$ state, so for a particular realisation of the thermal sampling the system wavefunction is

$$|\Psi_0\rangle \equiv \bigotimes_{j=1}^{N} |\phi_j, \downarrow\rangle_j, \tag{S7}$$

where $|\phi_j\rangle$ is the spatial state of the $j^{\text{th}}$ atom. The normalisation is chosen such that $\langle \Psi_0 | \Psi_0 \rangle = 1$, and therefore initially $\vec{S} = N(0, 0, -1/2)$.

Between the two RF pulses of the Ramsey sequence, the azimuthal angle $\varphi$ of the collective spin winds at a rate $\Omega(t) \equiv \partial \varphi(t') / \partial t'|_{t'=t}$. The contribution to $\Omega$ of two-body correlations, denoted $\Omega_2$, can be straightforwardly calculated at $t = 0$ by noting that after the first RF pulse $\vec{S} = N(0, -1/2, 0)$, and so

$$\Omega_2(0) = \frac{1}{|\vec{S}|} \frac{d\langle \hat{S}_x \rangle}{dt} = \frac{2i}{N\hbar} \langle \left[ \hat{H}, \hat{S}_x \right] \rangle, \tag{S8}$$

where $\hat{H}$ is the Hamiltonian for a Bose gas displaying $\uparrow\uparrow$ contact interactions characterised by scattering length $a$, and negligible $\uparrow\downarrow$ and $\downarrow\downarrow$ interactions. By adapting the analogous calculation for two-component Fermi gases in [9] the commutator of Eq. (S8) can be computed, yielding $\Omega_2(0) =$

$C_2 \frac{\hbar}{4\pi mn}(a^{-1} - r_0^{-1})$ where $C_2$ is the two-body contact density, $r_0$ the range of the interatomic potential and $n$ the number density of the interacting component.

We also note that Eq. (S8) is equivalent to the energy cost for an infinitesimal coherent vertical rotation of $\vec{S}$ on the Bloch sphere; the energy cost per photon for a rotation of angle $\delta\theta$ is $\hbar\Omega = (\langle e^{-i\delta\theta\hat{S}_x}\hat{H}e^{i\delta\theta\hat{S}_x}\rangle - \langle\hat{H}\rangle)/(N\delta\theta/2)$ which recovers Eq. (S8) in the limit of infinitesimal transfer $\delta\theta \to 0$. The precession frequency at $t = 0$ obtained via Ramsey spectroscopy is therefore identical to the shift in resonant frequency for RF transitions between the spin states.

Here we present a more complete approach, which both yields $\Omega(t)$ for any value of $t$, and incorporates the influence of the three-body correlations. First, we show that $S(t)$ is related by a Fourier transform to the RF excitation spectrum $\Gamma(\omega)$ for transferring atoms between $\downarrow$ and $\uparrow$. Second, we use the known $\Gamma(\omega)$ for a strongly-interacting Bose gas [10] to evaluate $\Omega(t)$.

## (1) Fourier relation between $S(t)$ and $\Gamma(\omega)$

The first RF pulse of the Ramsey sequence rotates the spin of each atom into a superposition of $\uparrow$ and $\downarrow$, without changing the spatial wavefunction. Subsequently, the state of the gas evolves as

$$|\Psi(t)\rangle = e^{-i\frac{t}{\hbar}\hat{H}}\left[\bigotimes_{j=1}^{N}\left(|\phi_j,\uparrow\rangle + |\phi_j,\downarrow\rangle\right)_j\right]2^{-N/2}, \quad (S9)$$

where $\hat{H}$ is the total Hamiltonian, which crucially does not couple different spin states. For $N \gg 1$ the binomial expansion of the $N$-body spin wavefunction product in Eq. (S9) is strongly dominated by terms corresponding to an approximately equal mixture of $|\uparrow\rangle$ and $|\downarrow\rangle$ atoms.

The time evolution of $|\Psi(t)\rangle$ leads to an evolution of the collective spin $\vec{S}$ in the equatorial plane of the Bloch sphere. We denote its position by a complex number

$$S(t) = |S(t)|e^{i\varphi(t)}, \quad (S10)$$

which is defined as

$$S(t) = \langle\hat{S}_x\rangle + i\langle\hat{S}_y\rangle = \langle\hat{S}_+\rangle$$
$$= \langle\Psi(t)|\sum_{j=1}^{N}\hat{s}_{j,+}|\Psi(t)\rangle \quad (S11)$$

where $\hat{s}_{j,+}$ is the spin-raising operator acting on the $j^{th}$ particle.

In evaluating Eq. (S11), the presence of single-particle operators means that it is convenient to separate out a single 'probe' atom from the interacting 'bath'. Since $N \gg 1$ the presence of this probe particle does not appreciably influence the evolution of the remaining $N - 1$ atoms, which constitute an interacting bath whose many-body state we denote $|\eta\rangle$. We separate the Hamiltonian into two terms,

$$\hat{H} \approx \hat{H}_\eta + \hat{H}_p(\eta), \quad (S12)$$

where $\hat{H}_\eta$ acts only on the atoms in the bath. The second term $\hat{H}_p(\eta)$ acts only on the probe atom, but is a function of the bath state $|\eta\rangle$ which may display complicated many-body dynamics. This separation amounts to the assumption that the probe atom and bath remain in a product state during the evolution, and hence neglects entanglement between them. The consequence is that the state of the system separates, giving

$$|\Psi(t)\rangle \approx \underbrace{\left[e^{-i\frac{t}{\hbar}\hat{H}_p(\eta)}\left(|\phi_1,\uparrow\rangle + |\phi_1,\downarrow\rangle\right)_1\right]}_{\text{State of probe atom}} \otimes \underbrace{\left[e^{-i\frac{t}{\hbar}\hat{H}_\eta}\left(\bigotimes_{j=2}^{\frac{N}{2}}\bigotimes_{k=\frac{N}{2}+1}^{N}|\phi_j,\uparrow\rangle_j|\phi_k,\downarrow\rangle_k + \dots\right)\right]}_{\text{State of bath, }|\eta(t)\rangle}2^{-N/2}, \quad (S13)$$

where '$\dots$' refers to $\mathcal{O}(2^N)$ further terms, the vast majority of which are very similar to the one shown, up to permutations of particle indices and relative fluctuations of order $1/\sqrt{N}$ in the number of $\uparrow$ atoms. In this large-$N$ limit, the bath state $|\eta(t)\rangle$ therefore evolves like an equal mixture of interacting and non-interacting atoms.

The collective spin $S(t)$ of Eq. (S11) can now be evaluated. For notational simplicity we drop the particle index, giving

$$S(t) \approx \frac{N}{2}\langle\phi,\uparrow|e^{i\frac{t}{\hbar}\hat{H}_p(\eta)}\hat{s}_+e^{-i\frac{t}{\hbar}\hat{H}_p(\eta)}|\phi,\downarrow\rangle$$
$$= \frac{N}{2}\langle\phi,\uparrow|e^{i\frac{t}{\hbar}\hat{H}_p(\eta)}\hat{s}_+|\phi,\downarrow\rangle e^{-i\frac{t}{\hbar}E_\downarrow}. \quad (S14)$$

Here we exploit the fact that $|\phi,\downarrow\rangle$ is non-interacting and so remains an eigenstate, with energy $E_\downarrow$ defined by $\hat{H}_p(\eta)|\phi,\downarrow\rangle = E_\downarrow|\phi,\downarrow\rangle$.

The Ramsey signal $S(t)$ therefore compares the evolution of one $\downarrow$ atom which is transferred to the interacting $\uparrow$ state immediately, to another which is transferred after a time $t$. The magnitude of $S(t)$ reveals the spatial overlap of the wavefunction in the two cases, and its argument their relative phase.

We can relate the experimentally observed Ramsey signal $S(t)$ to the frequency response by a Fourier transform. Inserting the identity operator $\sum_f |f\rangle\langle f|$, where $\{f\}$ is a complete

basis of eigenstates of $\hat{H}_p(\eta)$, yields

$$\int dt\, S(t)e^{i\omega t} \propto \int dt \sum_f |\langle f|\, \hat{s}_+\, |\phi, \downarrow\rangle|^2 e^{i(E_f - E_\downarrow)t/\hbar} e^{i\omega t}$$

$$= \sum_f |\langle f|\, \hat{s}_+\, |\phi, \downarrow\rangle|^2 \delta\left(\omega - \left(\frac{E_f - E_\downarrow}{\hbar}\right)\right)$$

$$\propto \Gamma(\omega) \qquad (S15)$$

where we assume that the bath state $|\eta(t)\rangle$, and hence $|f\rangle$, does not evolve appreciably over the integration time. Here $\Gamma(\omega)$ is the RF excitation spectrum for small vertical tilts of $\vec{S}$, corresponding to an infinitesimal transfer of atoms between $\downarrow$ and $\uparrow$.

The quantity $S(t)$ is therefore the Fourier transform of the RF excitation spectrum $\Gamma(\omega)$. This conclusion is similar to the one obtained for Ramsey spectroscopy of a single impurity immersed in a Fermi sea [11–13].

## (2) Calculation of $\Omega(t)$

By taking the time-derivative of Eq. (S10), assuming that $|S(t)|$ varies slowly, and employing Eq. (S15), we find

$$\Omega(t) = \frac{\int_{-\infty}^{\infty} d\omega\, \omega\Gamma(\omega)e^{i\omega t}}{\int_{-\infty}^{\infty} d\omega\, \Gamma(\omega)e^{i\omega t}}. \qquad (S16)$$

To calculate $\Omega$, we evaluate the integrals in Eq. (S16) using $\Gamma(\omega)$ derived in [10]. This gives

$$\Omega = \frac{\hbar}{4\pi m}\left(\frac{1}{na}C_2 + \frac{5.0\,\pi^2}{n}C_3\right), \qquad (S17)$$

which is the result quoted in the main paper. Here the contact densities $C_2$ and $C_3$ correspond to the correlations in an interacting Bose gas of number density $n$, which in our Ramsey experiment is equal to half the total density.

Explicitly, the calculations of the integrals in Eq. (S16) are

$$\int_{-\infty}^{\infty} d\omega\, \omega\Gamma(\omega)e^{i\omega t} = \Omega_{\mathrm{R}}^2 V \int_{-\infty}^{\infty} d\omega\, \frac{ie^{i\omega t}}{2}\left(\left[\frac{-1}{4\pi ma}\left(\frac{1}{\omega + i\varepsilon} - \frac{1}{\omega - i\varepsilon}\right) + \dots\right]C_2 - \frac{1}{m}\left[\frac{F_{\mathrm{rf}}(\omega + i\varepsilon)}{(\omega + i\varepsilon)} - \frac{F_{\mathrm{rf}}(\omega - i\varepsilon)}{(\omega - i\varepsilon)}\right]C_3\right)$$

$$= \Omega_{\mathrm{R}}^2 V \int_{-\infty}^{\infty} d\omega\, \frac{ie^{i\omega t}}{2}\left(\left[\frac{2\pi i}{4\pi ma}\delta(\omega) + \dots\right]C_2 + \left[\frac{D_0 + D_2/2}{m}2\pi i\delta(\omega) + \dots\right]C_3\right)$$

$$\approx -\Omega_{\mathrm{R}}^2 \pi V \frac{\hbar}{m}\left(\frac{1}{4\pi a}C_2 + 1.25\pi C_3\right) \qquad (S18)$$

$$\int_{-\infty}^{\infty} d\omega\, \Gamma(\omega)e^{i\omega t} = \Omega_{\mathrm{R}}^2 \int_{-\infty}^{\infty} d\omega\, \frac{ie^{i\omega t}}{2}\left(\left[\frac{1}{\omega + i\varepsilon} - \frac{1}{\omega - i\varepsilon}\right]N + \dots\right)$$

$$= \Omega_{\mathrm{R}}^2 \int_{-\infty}^{\infty} d\omega\, \frac{ie^{i\omega t}}{2}(-2\pi i\delta(\omega)N + \dots)$$

$$\approx \Omega_{\mathrm{R}}^2 \pi N \qquad (S19)$$

where $\Omega_{\mathrm{R}}$ is the Rabi frequency of the RF drive, $\varepsilon$ is an infinitesimal real number, $V$ is the system volume, and both the function $F_{\mathrm{rf}}$ and the constants $D_0$ and $D_2$ are defined in [10].

The '...' in Eqs. (S18, S19) denote the broad high-frequency contributions which dephase on a timescale $\sim mr_0^2/\hbar$. For $^{39}$K, where $r_0 = 64\, a_0$, this timescale is $\sim 10$ ns which is much shorter than the timescales of our experiment.

## Relationship of $\varphi$ and $\delta_0$

Initially, all atoms are in the $\downarrow$ state and $\vec{S} = N(0, 0, -1/2)$. During the Ramsey sequence, $\vec{S}$ acquires an azimuthal phase on the Bloch sphere both during the RF pulses of duration $t_{\mathrm{p}}$, and during the evolution time $T$ between the pulses. We consider each in turn.

Neglecting interactions during an RF pulse, the evolution of $\vec{S}$ in the presence of an RF drive is described by the optical Bloch equations [14],

$$\dot{\vec{S}} = \vec{S} \times \vec{W}. \qquad (S20)$$

Here $\vec{W} = (\Omega_{\mathrm{R}}, 0, \delta)$, where $\delta$ is the detuning from the non-interacting resonance. This equation describes precession of $\vec{S}$ about $\vec{W}$ at an angular frequency $\sqrt{\Omega_{\mathrm{R}}^2 + \delta^2}$.

In our case, $\Omega_{\mathrm{R}}t_{\mathrm{p}} = \pi/2$, and $\vec{S}$ precesses about $\vec{W}$ by an angle $(\pi/2)\sqrt{1 + (\delta/\Omega_{\mathrm{R}})^2}$. Assuming $\delta/\Omega_{\mathrm{R}} \ll 1$, to first order we find that immediately after the first pulse $\vec{S} = (\delta/\Omega_{\mathrm{R}}, -1, 0)$, giving a phase advance of $\delta/\Omega_{\mathrm{R}}$ relative to the case of a resonant RF drive.

Between the two RF pulses, $\vec{S}$ accumulates a phase $\varphi$ due to interactions, and a phase $T\delta$ due to the detuning of the RF

source. The total accumulated azimuthal phase $\Phi$ is therefore

$$\Phi = \underbrace{T\delta + \varphi}_{\substack{\text{Between the} \\ \text{RF pulses}}} + \underbrace{2\delta/\Omega_{\mathrm{R}} + \Delta\varphi_{\mathrm{p}}}_{\substack{\text{During the} \\ \text{RF pulses}}}, \qquad \text{(S21)}$$

where $\Delta\varphi_{\mathrm{p}}$ is the interaction-induced phase acquired during the pulses. The centre of the Ramsey fringes occurs at the detuning $\delta_0$ for which $\Phi = 0$, and from Eq. (S21) we find

$$\delta_0 = -\frac{\varphi + \Delta\varphi_{\mathrm{p}}}{T + 4t_{\mathrm{p}}/\pi}, \qquad \text{(S22)}$$

which reproduces Eq. (2) of the main paper.

The small $\Delta\varphi_{\mathrm{p}}$ is generally complicated to calculate. A major contribution in our experiment is an RF-induced shift of the position of the Feshbach resonance. This modifies the interaction strength, and hence $\Omega$, during the pulses [15, 16].

## SIMULATION OF TWO-BODY DYNAMICS FOLLOWING AN INTERACTION QUENCH

Before the first RF pulse, the gas is essentially non-interacting, meaning that correlations (and hence both contacts) are negligible. Since the pulse cannot transfer momentum to the atoms, it leaves the spatial wavefunction unchanged and correlations gradually develop during the evolution time after the quench. In this section we estimate the timescale for the development of two-body correlations using a simple model, in the spirit of [17].

We consider two atoms of mass $m$ at positions $\vec{r}_1$ and $\vec{r}_2$, interacting via a contact potential

$$V(r) = \frac{4\pi\hbar^2 a}{m}\delta(r), \qquad \text{(S23)}$$

where $r = |\vec{r}_2 - \vec{r}_1|$. To normalise the wavefunctions, atoms are confined to remain within a distance $R$ of each other; we have checked that none of our conclusions depend on the specific choice of $R$.

We assume that the centre-of-mass is untrapped, hence the absolute-position eigenstate is simply a plane wave which we take to be zero-momentum. On the other hand, the relative-motion eigenstates $|n\rangle$ are

$$\langle r|n\rangle = B_n \frac{\sin(k_n r + \delta(k_n))}{r}, \qquad \text{(S24)}$$

where $B_n$ is a normalisation factor, and the wavevectors $k_n$ are determined by the boundary condition

$$k_n R + \delta(k_n) = n\pi, \qquad n \in \mathbb{Z}. \qquad \text{(S25)}$$

We work in the zero-range limit such that the scattering phase shift is $\delta(k) = -\arctan(ka)$, and only consider $a < 0$ such that the scattering states $|n\rangle$ form a complete basis.

Immediately before the quench at time $t = 0$ atoms are in a non-interacting state $|i\rangle$,

$$\langle r|i\rangle = \frac{1}{\sqrt{2\pi R}}\frac{\sin k_0 r}{r}, \qquad \text{(S26)}$$

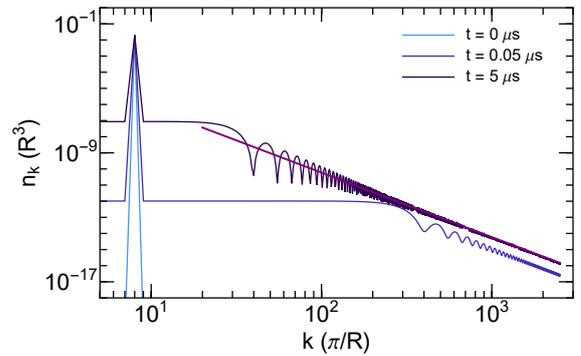

FIG. S3. **Post-quench momentum dynamics.** The momentum distribution $n_k$ at several times $t$ after a quench of the interaction strength from zero to $\lambda/a = -25$. As time progresses a high-momentum $k^{-4}$ tail develops, indicating the evolution of two-body correlations. The solid purple line denotes the momentum distribution corresponding to the theoretical equilibrium value for $C_2$. Note that only data corresponding to wavevectors which satisfy $k = n\pi/R$, where $n$ is an integer, are plotted.

with some initial relative wavevector $k_0$. We use a characteristic value $k_0 = \sqrt{8}/\lambda_0$, where $\lambda_0 = 8600\, a_0$ is the thermal wavelength of the gas used in our experiments.

At a time $t$ after the quench, the system state $|\psi(t)\rangle$ is given by

$$|\psi(t)\rangle = \sum_n \langle n|i\rangle \; e^{-i\varepsilon_n t/\hbar} \; |n\rangle, \qquad \text{(S27)}$$

where $\varepsilon_n = \hbar^2 k_n^2/(2\mu)$, and $\mu = m/2$ is the reduced mass of the particle pair.

We are interested in the evolution of the momentum distribution $n_k(t)$, the high-momentum part of which reflects the short-distance behaviour of the spatial wavefunction, and hence the strength of two-body correlations. Denoting a relative-momentum state $|k\rangle$, the momentum distribution is computed as

$$\begin{aligned} n_k(t) &= |\langle k|\psi(t)\rangle|^2, \\ &= \left| \sum_n \langle n|i\rangle \; e^{-i\varepsilon_n t/\hbar} \; \langle k|n\rangle \right|^2, \end{aligned} \qquad \text{(S28)}$$

where the momentum-space state $\langle k|n\rangle$ is the Fourier transform of $\langle r|n\rangle$ defined in Eq. (S24). The normalisation of $\langle k|n\rangle$ is chosen such that $\int \frac{\mathrm{d}^3 k}{(2\pi)^3}\, n_k = N$, where $N = 2$ is the number of particles. Since the centre-of-mass of the particle pair is untrapped, the relative-momentum distribution and single-particle momentum distribution coincide.

In Fig. S3 we show the evolution of $n_k(t)$ after an interaction quench to $\lambda_0/a = -25$. It gradually develops a $k^{-4}$ tail at high momenta, corresponding to the expected $1/r$ variation of the spatial wavefunction at small particle separations. The two body contact density is defined such that

$$\lim_{k\to\infty} n_k = \frac{C_2 V}{k^4}, \qquad \text{(S29)}$$

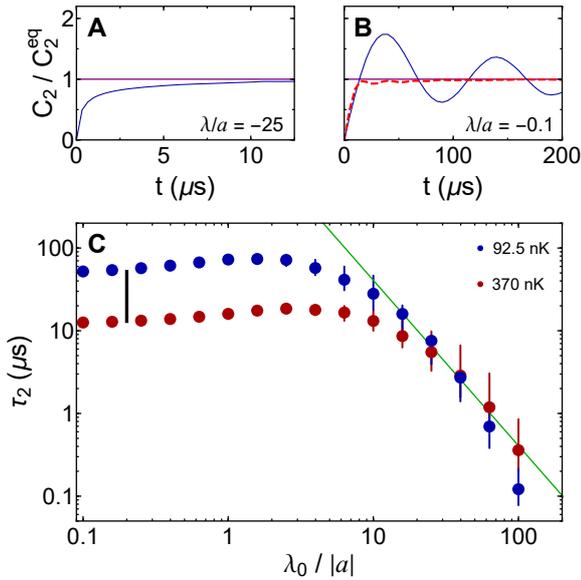

FIG. S4. **Timescale for the development of two-body correlations.** (**A**) $C_2(t)$ for an atom pair quenched to weak interactions (blue line), which damps towards its equilibrium value (purple line). (**B**) In the case of a quench to the unitary regime, oscillations in $C_2$ are observed. The red dashed line indicates the average $C_2(t)$ obtained for atoms pairs with a range of relative momenta. (**C**) The characteristic growth timescale $\tau_2$ extracted from the $C_2(t)$ curves (see text). Timescales are obtained for atom pairs with a characteristic temperature of 370 nK (red) and 92.5 nK (blue). The solid green line denotes $\tau_2 \propto a^2$ to guide the eye, and the black vertical bar denotes the expected ratio of 4 between the saturation plateaus for the two data sets.

where $V$ is the system volume, and we therefore extract $C_2$ by a $k^{-4}$ fit to $n_k$ at high momenta. In addition to the dominant $k^{-4}$ behaviour, $n_k$ exhibits oscillations; these are a consequence of non-equilibrium dynamics induced by the quench, and similar observations were reported in [17].

In Fig. S4 we show $C_2$ against time for a pair of atoms quenched to: (A) $\lambda_0/a = -25$ and (B) $\lambda_0/a = -0.1$. In both cases $C_2$ is normalised to the corresponding equilibrium contact $C_2^{\text{eq}}$, which is calculated below. Remarkably, despite the purely unitary evolution of the wavefunctions, $C_2/C_2^{\text{eq}}$ damps to unity at long times.

We find qualitatively different behaviour of $C_2(t)$ for unitary and non-unitary interaction strengths. For weak interactions $C_2$ monotonically approaches $C_2^{\text{eq}}$. On the other hand, for unitary interactions it displays oscillations which gradually damp towards $C_2^{\text{eq}}$. The period of oscillation depends on the relative wavevector, and in a thermal ensemble we expect the oscillations from atom pairs with different relative momenta to dephase. The blue line in Fig. S4B shows the behaviour obtained for a single atom pair with relative wavevector $k_0$, and the red dashed line shows the average of all curves obtained for relative wavevectors $k_0 \times \{0.05, 0.1 \ldots 2\}$.

To obtain a characteristic timescale for the growth of $C_2$, at all interactions strengths we extract the time $\tau_2$ at which $C_2$ has grown to 90% of its equilibrium value. This timescale is plotted in Fig. S4C, for simulations in which $k_0 = \sqrt{8}/\lambda_0$ (red, corresponding to the temperature of 370 nK used in our Ramsey experiments), and additionally for $k_0 = \sqrt{2}/\lambda_0$ (blue, corresponding to a temperature of 92.5 nK) for comparison.

Away from unitarity, the two curves coincide and show the expected $a^2$ scaling that reflects the two-body energy scale; the green line shows $\tau_2 = 30\, ma^2/\hbar$. At unitarity, $\tau_2$ saturates to a value limited by the cloud temperature. The two data sets saturate to different values, with the vertical black bar denoting the expected ratio of 4 between them. The value of $\tau_2$ in the unitary plateaus is approximately $0.1\, m\lambda^2/\hbar$.

For the red data, which corresponds to the temperature of the gas used in our experiments, $\tau_2 < 20\,\mu$s for all interaction strengths. Since the duration of our RF pulse is 17 $\mu$s, and the shortest interrogation time is 30 $\mu$s, we do not expect to resolve any dynamics of $C_2$.

*Calculation of the equilibrium contact* We calculate $C_2^{\text{eq}}$ using the adiabatic relation,

$$C_2^{\text{eq}} = \frac{8\pi m a^2}{\hbar^2 V}\left(\frac{\partial E}{\partial a}\right).\tag{S30}$$

We consider two atoms in an eigenstate with relative momentum $k$, which satisfies the boundary condition of Eq. (S25), $kR + \delta(k) = m\pi$, where $m$ is an integer. Their energy at a non-zero interaction strength, relative to the non-interacting case, is

$$\Delta E = \frac{\hbar^2}{2\mu}\left(k^2 - \left(\frac{m\pi}{R}\right)^2\right)\tag{S31}$$

$$= \frac{\hbar^2}{2\mu R^2}\left(\delta^2 - 2m\pi\delta\right).\tag{S32}$$

The equilibrium contact density $C_2^{\text{eq}}$ for the two atoms can now be evaluated according to Eq. (S30),

$$C_2^{\text{eq}} = \frac{8\pi m a^2}{V\hbar^2}\left(\frac{\partial E}{\partial a}\right)\tag{S33}$$

$$= \frac{8\pi m a^2}{V\hbar^2}\frac{\hbar^2}{\mu R^2}\left(\delta - m\pi\right)\frac{-k}{1 + (ka)^2}\tag{S34}$$

$$= \frac{12}{R^4}\frac{(ka)^2}{1 + (ka)^2}.\tag{S35}$$

This result agrees very well with the long-time limit of the numerically-obtained $C_2(t)$ curves (see Fig. S4).



\end{document}